\documentclass[journal]{IEEEtran}

\usepackage[compatibility=false]{caption}
\ifCLASSINFOpdf
\else
   \usepackage[dvips]{graphicx}
\fi
\usepackage{url}
\usepackage{graphicx}
\usepackage{amsmath}
\usepackage{subcaption}
\usepackage{multirow}
\usepackage{booktabs}
\usepackage{makecell}
\usepackage[compatibility=false]{caption}
\usepackage{cite}
\usepackage{float}

\begin{document}

\title{Pinhole Effect on Linkability and Dispersion in Speaker Anonymization}

\author{Kong Aik Lee, Zeyan Liu, Liping Chen, and Zhenhua Ling
\thanks{Kong Aik Lee is with The Hong Kong Polytechnic University, Hong Kong (e-mail: kong-aik.lee@polyu.edu.hk). Zeyan Liu, Liping Chen, and Zhenhua Ling are with the University of Science and Technology of China, China (e-mail: xy671231@mail.ustc.edu.cn, lipchen@ustc.edu.cn, zhling@ustc.edu.cn).}
\thanks{\emph{Corresponding author: Kong Aik Lee.}}}

\markboth{Journal of \LaTeX\ Class Files, Vol. 14, No. 8, August 2015}
{Shell \MakeLowercase{\textit{et al.}}: Bare Demo of IEEEtran.cls for IEEE Journals}
\maketitle

\markboth{}{}
\thispagestyle{plain}
\pagestyle{plain}

\begin{abstract}
Speaker anonymization aims to conceal speaker-specific attributes in speech signals, making the anonymized speech unlinkable to the original speaker identity. Recent approaches achieve this by disentangling speech into content and speaker components, replacing the latter with pseudo speakers. The anonymized speech can be mapped either to a common pseudo speaker shared across utterances or to distinct pseudo speakers unique to each utterance. This paper investigates the impact of these mapping strategies on three key dimensions: speaker linkability, dispersion in the anonymized speaker space, and de-identification from the original identity. Our findings show that using distinct pseudo speakers increases speaker dispersion and reduces linkability compared to common pseudo-speaker mapping, thereby enhancing privacy preservation. These observations are interpreted through the proposed \emph{pinhole effect}, a conceptual framework introduced to explain the relationship between mapping strategies and anonymization performance. The hypothesis is validated through empirical evaluation.

\end{abstract}



\IEEEpeerreviewmaketitle

\section{Introduction}

\IEEEPARstart{S}{peaker} anonymization is the task of altering the speaker’s voice to hide their identity to the greatest possible extent (i.e., to human perception and machine recognition), while leaving all other speech attributes intact \cite{Tomashenko2022VoicePrivacy2020, Panariello2024VoicePrivacy2022, Chen2025PseudoSpeaker}. For instance, speech signals are anonymized to conceal the identity of the interviewee on a television program while keeping the spoken contents. In a wider context, speaker anonymization is posed as a privacy-preservation solution, alongside homomorphic encryption \cite{Pathak2013PrivacyPreserving, Zhang2019EncryptedSpeech} and federated learning \cite{Leroy2019Federated}. Different from the latter, speaker anonymization transforms speech signals into a privacy-preserving format that aligns with current pipelines. This practicality has led to its widespread adoption, as anonymized speech data can be used in downstream speech processing tasks (e.g., speech and emotion recognition) with minimal or no modifications to existing systems, while preserving speaker’s privacy.


Mainstream speaker anonymization approach is based on replacing the speaker’s voice attributes with those of a pseudo speaker \cite{Fang2019SpeakerAnonymization, Srivastava2022XvectorAnonymization, Chen2025PseudoSpeaker}. In this approach, the  input speech could be anonymized to a common or unique pseudo-speaker. Both of these have their pros and cons. For instance, unique pseudo speakers are useful for a multi-party conversational setting when multiple speakers are involved. In contrast, a common pseudo-speaker is commonly used in a passer-by interview. It has been stipulated that mapping to a common pseudo speaker would lead to more effective privacy preservation since all anonymized speech would sound alike. This paper shows that this first intuition does not hold. More specifically, we hypothesize that mapping to distinct pseudo-speakers reduces linkability (i.e., likelihood of re-identification) and thereby enhances privacy preservation. We validate this hypothesis through experiments using two different speaker anonymization systems. Furthermore, we demonstrate that this phenomenon can be explained by the \emph{pinhole effect}, a conceptual framework proposed for the first time in this paper.

\begin{figure*}[t!]
    \begin{subfigure}[t]{0.5\textwidth}
        \centering
        \includegraphics[width=0.8\columnwidth]{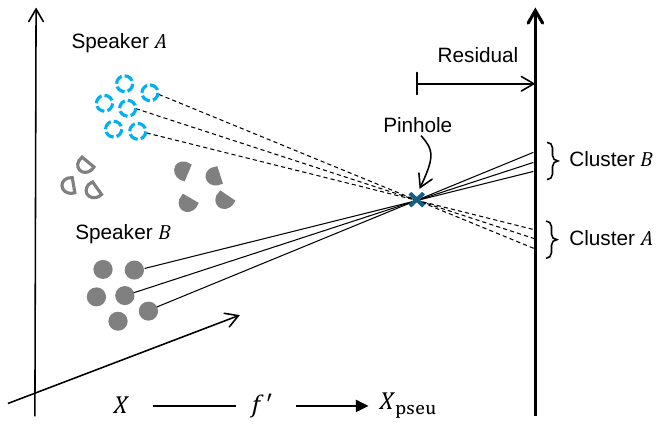}
        \caption{}
    \end{subfigure}
    \begin{subfigure}[t]{0.5\textwidth}
        \centering
        \includegraphics[width=0.8\columnwidth]{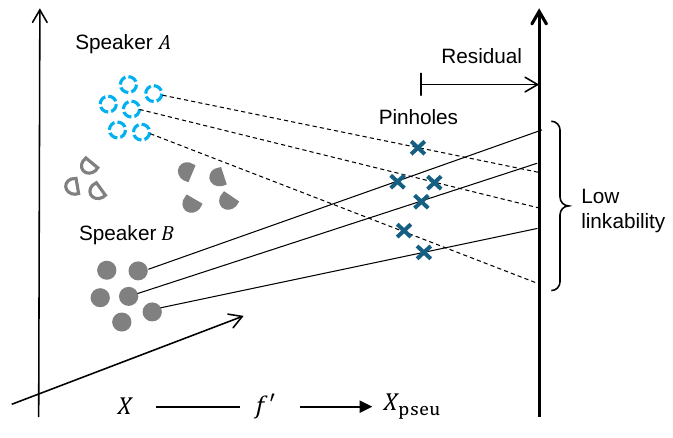}
        \caption{}
    \end{subfigure}
    \caption{Pinhole effect in speaker anonymization: (a) a common pseudo-speaker is used across all anonymization instances, and (b) distinct pseudo-speaker for each anonymized utterances.}
    \label{fig:pinhole-effect}
\end{figure*}

\section{Speaker Anonymization}

Speaker anonymization belongs to a subclass of privacy-preserving technology (PPT). PPT aims to protect data privacy while the data is being processed, at rest on a system, or in transit between systems \cite{archer2023}. Put it formally. Let $X_p$ be the private data in $X$, where $X_p$ leads to the inference of $Y$ (e.g., identity of the person, gender, ethnicity):
\begin{equation}
    X_p\rightarrow Y
\end{equation}
Privacy preservation aims to keep private data $X_p$ opaque to an insider or an outsider, when the data $X$ is being processed, in transit, or stored. 

Speaker anonymization realizes the goal of privacy preservation by removing or concealing the voice attributes $X_v\subset X_p$ that leads to the inference of speaker identity $Y_\text{ID}\in Y$, i.e., $X_v\rightarrow Y_\text{ID}$, while keeping other private and non-private information (linguistic and para-linguistic) untouched. Formally, let $X$ be a speech signal, speaker anonymization encompasses a mapping between the input and the anonymized speech, represented by the function $f$, as follows:
\begin{equation}
    f:X\longmapsto X\backslash X_v
\end{equation}
where the shorthand $X\backslash X_v$ denotes the set $X$ excluding the subset $X_v$. The resulting speech signal $X\backslash X_v$ is referred to as the anonymized speech. In practice, since a speech signal cannot exist without a speaker’s voice as a carrier of the spoken words, the mapping is realized as 
\begin{equation}
    f^\prime:X\longmapsto(X\backslash X_v)\cup X_\text{pseu}
\end{equation}
where $X_\text{pseu}$ represents the pseudo-speaker voice introduced to replace $X_v$. Pseudo-speakers are artificial speakers created by algorithms that are not linked to any real person, thereby avoiding infringing the privacy of real individuals. Many approaches have been proposed for the generation and selection of pseudo speakers \cite{Chen2025PseudoSpeaker, Srivastava2022XvectorAnonymization, Champion2022,Miao2023,Yao2024,Turner2022,Meyer2023}. For example, by taking the average of a subset of speakers selected from a cohort set with the further distance away from the source speaker \cite{Srivastava2020}.  

\section{Pinhole Effect in Speaker Anonymization}

In the anonymization process, an important consideration is whether to use the same pseudo-speaker across different anonymization instances. One approach is to apply a universal pseudo-speaker to all anonymized utterances, while another is to assign a unique pseudo-speaker to each utterance. The former leads to an any-to-one mapping, while the latter leads to an any-to-any mapping. 

The mapping process can be shown more concretely using (3). In the case when a universal voice $X_\text{pseu}^i=X_\text{pseu}$ is used for all $i$ anonymization instances, the anonymization function $f^\prime$ becomes an any-to-one mapping. On the other hand, the function $f^\prime$ implements an any-to-any mapping when $X_\text{pseu}^i\neq X_\text{pseu}^j$ for $i\neq j$. 
In Figs.~\ref{fig:pinhole-effect} (a) and (b), we illustrate the any-to-one and any-to-any mappings graphically assuming a two-dimensional feature space. Due to the imperfection in the removal of speaker voice attributes, residual attributes cause the anonymized utterances from the same speaker to cluster together. In Fig.~\ref{fig:pinhole-effect}(a), this is illustrated as beams of light passing through a pinhole, where beams originating from the same source (i.e., in analogy to speech utterances from the same speaker) passing through the pinhole will cluster around in the same area. In Fig.~\ref{fig:pinhole-effect}(b), by using multiple pinholes (i.e., analogous to multiple pseudo-voices), the anonymized utterances from the same speakers are scattered apart in the anonymized space. In this case, the pseudo-voice attributes overwhelm the residual attributes of the original speakers. 

The \emph{pinhole effect} and its implications for speaker anonymization, as illustrated in Fig.~\ref{fig:pinhole-effect}, can be summarized as follows:
\begin{itemize}
    \item \textbf{Dispersion:} Any-to-any mapping leads to greater dispersion in speaker representations of anonymized speech compared to any-to-one mapping.
    \item \textbf{Linkability:} Any-to-any mapping reduces speaker similarity among anonymized utterances, thereby lowering linkability relative to any-to-one mapping.
    \item \textbf{De-identification:} The speaker similarity between original and anonymized speech does not differ significantly regardless of the number of pinholes (one or multiple). Thus, both any-to-one and any-to-any mappings achieve a comparable level of de-identification.
\end{itemize}
\noindent The implication for \emph{speaker linkability} arises directly from the assertion on \emph{speaker dispersion}, as greater dispersion leads to reduced linkablity between anonymized utterances. Both of these relate to the distribution of anonymized speaker representations on the right-hand side of the pinhole(s) in Figs.~\ref{fig:pinhole-effect} (a) and (b). In contrast, the assertion concerning speaker \emph{de-identification} involves a cross-side comparison, evaluating the speaker similarity between original speech (left side) and anonymized speech (right side) across the pinhole(s).

\if 

\begin{figure*}[t!]
    \begin{subfigure}[t]{0.5\textwidth}
        \centering
        \includegraphics[width=0.7\columnwidth]{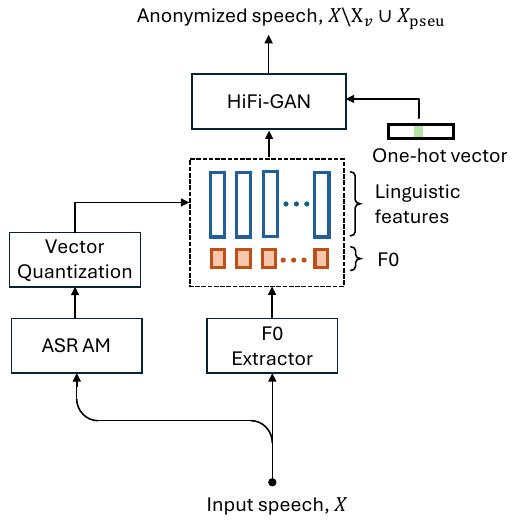}
        \caption{\texttt{SYS1}: One-hot speaker representation.}
        \label{fig:speaker-anonymization-a}
    \end{subfigure}
    \begin{subfigure}[t]{0.5\textwidth}
        \centering
        \includegraphics[width=0.7\columnwidth]{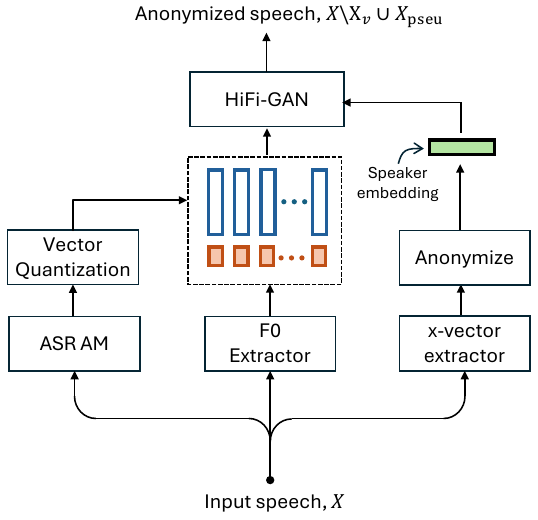}
        \caption{\texttt{SYS2}: Continuous speaker embedding.}
        \label{fig:speaker-anonymization-b}
    \end{subfigure}
    \caption{Two speaker anonymization systems used in the experiments to validate the \emph{pinhole effect}.}
    \label{fig:speaker-anonymization}
\end{figure*}

\else

\begin{figure}[t!]
    \centering
    \subfloat[One-hot speaker representation.]{\includegraphics[width=0.3\textwidth]{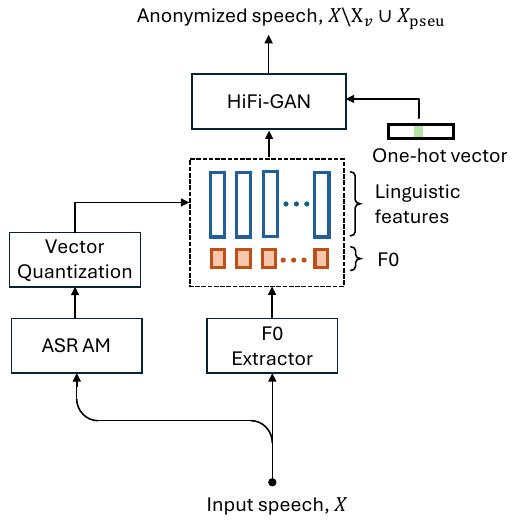}\label{fig:one-hot}}
    \hfill
    \subfloat[Continuous speaker embedding.]{\includegraphics[width=0.3\textwidth]{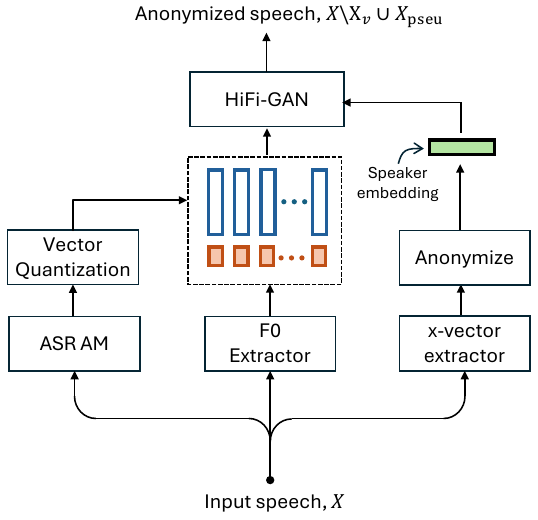}\label{fig:speaker-embedding}}
    \caption{Two speaker anonymization systems used in the experiments to validate the \emph{pinhole effect}.}
    \label{fig:speaker-anonymization}
\end{figure}

\fi


\section{Experiment}
We validate the pinhole effect described in Section III using two speaker anonymization systems developed for the VoicePrivacy 2024 (VPC2024) Challenge \cite{vpc2024}. The systems are evaluated under two anonymization settings. In the first setting, all utterances are anonymized to a common universal voice, corresponding to the any-to-one mapping strategy. In the second setting, each utterance is anonymized using a different pseudo voice, corresponding to the any-to-any mapping strategy.

\subsection{Models}

We used two speaker anonymization systems in the experiments. The first system (\texttt{SYS1}) corresponds to the B5 baseline in VPC2024. As depicted in Fig.~\ref{fig:speaker-anonymization}(a), the system consists of (1) an ASR acoustic model (ASR AM) to extract speech features containing the linguistic content, and (2) a pitch tracking model to extract F0 features. Vector quantization (VQ) is applied to the linguistic features, for which continuous feature vectors are replaced with discrete tokens. The quantization process introduces an information bottleneck, thereby reducing the residual speaker's voice attributes. The quantized features are then fed together with the $F0$ prosodic feature into a HiFi-GAN neural source-filter model~\cite{kong2020hifi} to synthesize an anonymized speech. \texttt{SYS1} can be configured for either any-to-one or any-to-any mapping by selecting an appropriate one-hot speaker vector. For the any-to-one mapping, a constant one-hot speaker ID is used for all anonymized utterances to produce a common universal voice. For the any-to-any mapping, a different one-hot speaker ID is randomly assigned to each anonymized utterance. In the any-to-one setting, speaker ID 103 was arbitrarily chosen from a pool of 251 speakers for use in the experiments.



The second system (\texttt{SYS2}) is a variant of the B5 baseline, where the one-hot vector input to the HiFi-GAN vocoder is replaced with a speaker embedding. As shown in Fig.~\ref{fig:speaker-anonymization}(b), the pseudo-speaker embedding $\phi_\text{pseu}$ is fed to a vocoder with the other two components (i.e., the $F0$ and linguistic features) to synthesize the anonymized speech. \texttt{SYS2} can also be configured for either any-to-one or any-to-any mapping by manipulating the speaker embedding. In the experiments, utterances are anonymized to a common pseudo-speaker, obtained by averaging the x-vector embeddings from the \emph{LibriSpeech train-clean-100}~\cite{panayotov2015librispeech}. For any-to-any mapping, each utterance is anonymized with a unique pseudo-speaker, obtained by taking the average of $100$ x-vector embeddings randomly selected from \emph{LibriSpeech train-clean-100}. 

\subsection{Datasets}
The configurations and datasets used to train each module of the anonymization systems are described as follow. The ASR AM comprises a wav2vec2 \cite{wav2vec2} front-end with three additional TDNN-F layers \cite{povey18_interspeech}. As in the VPC2024 B5 baseline, the wav2vec2 model was pre-trained on \emph{VoxPopuli} dataset \cite{voxpopuli} and then fine-tuned on \emph{LibriSpeech train-clean-100} \cite{panayotov2015librispeech}. The VQ has a codebook size of $48$, with a dimensionality of $256$. The x-vector extractor \cite{snyder17_interspeech} was trained on the \emph{VoxCelb-1} and \emph{VoxCeleb-2} datasets \cite{voxceleb}. As in most implementations, the x-vector has a dimensionality of $512$. The HiFi-GAN vocoder was trained on the \emph{LibriSpeech train-clean-100} with both ASR-AM and x-vector extractor frozen. A Pytorch implementation of YAAPT pitch tracking algorithm \cite{kasi2002pitch} was used for The $F0$ extractor.  

\subsection{Performance metrics}
Our experiments were carried out following the evaluation protocol provided in VPC2024 \cite{vpc2024}. The privacy-preserving capability of the anonymization systems was evaluated using \emph{automatic speaker verification} (ASV) tests, with performance measured by the \emph{equal error rate} (EER). To minimize inference risk, anonymized speech should not be successfully verified by an ASV system, which corresponds to a higher EER. The ability to preserve linguistic content was assessed using \emph{automatic speech recognition} (ASR) tests, measured by the \emph{word error rate} (WER). Since the goal is to retain speech information other than the speaker’s identity, the anonymized speech should have a WER as close as possible to that of the original speech. 

\begin{table}
    \centering
    \caption{Evaluation of anonymization methods on the LibriSpeech dataset using ASV EER (where higher values indicate better anonymization) and ASR WER (where lower values indicate better linguistic preservation).}
    \label{tab:baseline}
        \begin{tabular}{c|c|c|c|c|c}
            \Xhline{1px}
            \multirow{1}{*}{} & 
            \multirow{1}{*}{Test set} & 
            \multirow{1}{*}{Partition} & 
            \multirow{1}{*}{\texttt{ORG}} & 
            \multirow{1}{*}{\texttt{SYS1}} & 
            \multirow{1}{*}{\texttt{SYS2}} \\
            \hline
            \multirow{5}{*}{EER (\%)} 
            & \multirow{2}{*}{libri-dev} 
            & F & 10.51 & 33.37 & 34.94 \\
            \cline{3-6}
            & & M & 0.93 & 31.94 & 34.32 \\
            \cline{2-6}
            & \multirow{2}{*}{libri-test} 
            & F & 8.76 & 31.84 & 33.73 \\
            \cline{3-6}
            & & M & 0.42 & 32.19 & 32.74 \\
            \cline{2-6}
            & \multicolumn{2}{c|}{avg} 
            & 5.16 & 32.23 & 33.93 \\
            \hline
            \multirow{2}{*}{WER (\%)} 
            & libri-dev & - 
            & 1.79 & 3.95 & 3.88\\
            \cline{2-6}
            & libri-test & - 
            & 1.84 & 4.15 & 4.01 \\
            \Xhline{1px}
        \end{tabular}
\end{table}

Table \ref{tab:baseline} shows the baseline performance of the two anonymization systems, \texttt{SYS1} and \texttt{SYS2}, using the any-to-one mapping to a universal voice. The ASV evaluations were conducted in a a gender-dependent manner. The average EERs across the four subsets are included. Comparing to the EERs in the \texttt{ORG} column, when no anonymization was applied, both anonymization systems increase tremendously the EER to over $30\%$. Both anonymization systems obtained WERs comparable with the original speech (\texttt{ORG}) without anonymization. These results show that the anonymization systems achieve good privacy preservation while retaining the linguistic information for downstream tasks. 

\begin{table}
\centering
\caption{Scatter values and ratio $J$ for original and anonymized speech under different systems and mapping strategies.}
\begin{tabular}{c|c|c|c|c}
\Xhline{1pt}
  Method & Mapping & $\text{Tr}(W^\top S_w W)$ & $\text{Tr}(W^\top S_b W)$ & $J$ \\
\hline
\texttt{ORG} & -- & 206.71 & 305.39 & 1.477 \\
\hline
\texttt{SYS1} & a2o & 674.27 & 30.14 & 0.047 \\
\hline
\texttt{SYS1} & a2a & 1224.04 & 38.19 & 0.031 \\
\hline
\texttt{SYS2} & a2o & 730.91 & 31.83 & 0.045 \\
\hline
\texttt{SYS2} & a2a & 2192.49 & 48.95 & 0.023 \\
\Xhline{1pt}
\end{tabular}
\label{tab:dispersion}
\end{table}

\subsection{Results}
In the first experiment, we examine the speaker dispersion of anonymized speech. Using the x-vector extractor described in Section-IV(A), a total of 28,539 utterances from 251 speakers in the \emph{LibriSpeech train-clean-100} dataset were represented as x-vector embeddings. The within-class and between-class scatter matrices, denoted as $S_w$ and $S_b$ respectively, were computed using the speaker labels. The trace of the within-class scatter matrix, $\text{Tr}(W^\top S_w W)$, quantifies the speaker class compactness. A smaller value indicates tighter clustering. In contrast, the trace of the between-class scatter matrix, $\text{Tr}(W^\top S_b W)$, measures the separation between speakers, where a higher value indicates better speaker separation. Here, $W$ is the matrix of eigenvectors of $S_w^{-1}S_b$.
The within-class and between-class scatter values for the original utterances (\texttt{ORG}) are reported in the first row of Table~\ref{tab:dispersion}. The last column shows the scatter ratio $J = \text{Tr}((W^\top S_w W)^{-1}(W^\top S_b W))$.
To evaluate the effect of anonymization, utterances were processed using \texttt{SYS1} and \texttt{SYS2}, and x-vector embeddings were extracted from the anonymized speech. While the between-class scatter reduces, the within-class scatter increases substantially after anonymization. The resulting reduction in the scatter ratio $J$ suggests a decrease in speaker similarity, and therefore, lower linkability among anonymized utterances from the same speaker. Comparing the mapping strategies, the any-to-any (a2a) mapping results in significantly higher dispersion than the any-to-one (a2o) mapping in both systems.

\begin{table}[t]
   \centering
   \caption{ASV EER (\%) on the VPC2024 LibriSpeech Dev and Test subsets under any-to-one (a2o) and any-to-any (a2a) mapping strategies.}
   \begin{tabular}{c|c|cc|cc|c}
        \Xhline{1pt}
        \multirow{2}{*}{Method} & 
        \multirow{2}{*}{Mapping} & 
        \multicolumn{2}{c|}{Libri-Dev} & 
        \multicolumn{2}{c|}{Libri-Test} & 
        \multirow{2}{*}{Avg} \\
        \cline{3-6}
        & & F & M & F & M & \\
        \hline
        \texttt{SYS1} & a2o & 33.37 & 31.94 & 31.84 & 32.19 & 32.23 \\
        \hline
        \texttt{SYS1} & a2a & 34.88 & 36.21 & 33.12 & 32.43 & 34.16 \\        
        \hline
        \texttt{SYS2} & a2o & 34.94 & 34.32 & 33.73 & 32.74 & 33.93 \\
        \hline
        \texttt{SYS2} & a2a & 37.03 & 35.84 & 34.37 & 36.62 & 35.97 \\
        \Xhline{1pt}
    \end{tabular}
    \label{tab:linkability}
\end{table}

In the second experiment, we examine the \emph{identity linkability} property asserted by the pinhole effect. We assume that an attacker attempts to verify the speaker identity using anonymized speech utterances. In this case, anonymized utterances were used for enrollment and test in ASV. We experimented with two settings. In the first setting utterances were anonymized using a common pseudo-speaker leading to a any-to-one mapping as in Fig \ref{fig:pinhole-effect}(a). In the second setting, utterances were anonymized with different pseudo speakers leading to a any-to-any mapping as in Fig \ref{fig:pinhole-effect}(b).
Table \ref{tab:linkability} shows the ASV EER on the VPC2024 LibriSpeech Dev and Test subsets. Looking at the second and third row for anonymization system \texttt{SYS1}, any-to-any mapping leads to higher EER compared to any-to-one mapping. The EER increment amount to $5.35\%$ on average. Similar trend is observed for anonymization system \texttt{SYS2} in the last two rows. The EER increment amount to $5.65\%$ on average when any-to-any mapping was used. These results validate the assertion that anonymization with unique pseudo speakers reduces the linkability between anonymized utterances compared to the case when a common pseudo speaker is used for all utterances.

\begin{table}[t]
   \centering
   \caption{ASV EER (\%) when original speech is used for enrollment and anonymized speech for testing, measuring de-identification.}
   \begin{tabular}{c|c|cc|cc|c}
        \Xhline{1pt}
        \multirow{2}{*}{Method} & 
        \multirow{2}{*}{Mapping} & 
        \multicolumn{2}{c|}{Libri-Dev} & 
        \multicolumn{2}{c|}{Libri-Test} & 
        \multirow{2}{*}{Avg} \\
        \cline{3-6}
        & & F & M & F & M & \\
        \hline
        \texttt{SYS1} & a2o & 47.87 & 49.38 & 50.34 & 48.80 & 49.10\\
        \hline
        \texttt{SYS1} & a2a & 47.58	& 48.27	& 48.72 & 51.00 & 48.89\\   
        \hline
        \texttt{SYS2} & a2o & 48.72 & 48.27 & 47.81 & 49.00	& 48.45\\
        \hline
        \texttt{SYS2} & a2a & 49.01 & 47.98 & 49.26 & 48.60 & 48.71\\
        \Xhline{1pt}
    \end{tabular}
    \label{tab:deidentification}
\end{table}

The third experiment examines the \emph{de-identification} property asserted by the pinhole effect. With reference to Fig.~\ref{fig:pinhole-effect} (a) and (b), this setting corresponds to the comparison of the original speech on the left to the anonymized speech on the right of the pinhole(s). In this scenario, we assume that an attacker attempts to verify an anonymized speech utterance as if it was spoken by the same speaker given an original speech utterance without anonymization. Table \ref{tab:deidentification} shows the ASV EER. The EERs are relatively higher compared to the EERs in Table \ref{tab:linkability}, since the enrollment utterances were not anonymized. Comparing the EERs for any-to-one and any-to-any mappings, there is no substantial differences between the use of universal voice or unique voices in anonymization. This observation validation the second assertion of the pinhole effect.         



\subsection{Statistical significance test}
We conducted bootstrap resampling to estimate the $95$\% confidence intervals (CIs) of the EER differences between a2a and a2o, denoted as $d_{\text{EER}}$, for each test condition in Table III. The detailed procedure is described in the Appendix. As shown in Table~\ref{tab:boot1}, the null hypothesis value (zero difference) falls outside the $95$\% CI, indicating that the EER differences are statistically significant ($p < 0.05$).

\begin{table}[t]
\caption{Bootstrap-estimated 95\% confidence intervals (CIs) of EER differences ($d_{\text{EER}}$) for four test conditions.}
\label{tab:boot1}
\resizebox{\columnwidth}{!}{
\begin{tabular}{lcccc}
\toprule
\textbf{System} & \textbf{Libri-Dev (F)} & \textbf{Libri-Dev (M)} & \textbf{Libri-Test (F)} & \textbf{Libri-Test (M)} \\
\midrule
\textbf{SYS1 (Mean)} & 1.75 & 3.17 & 1.77 & 2.55 \\
\textbf{95\% CI}     & 0.27--2.61 & 0.15--6.05 & 0.45--4.31 & 0.52--6.50 \\
\textbf{SYS2 (Mean)} & 2.28 & 3.42 & 2.11 & 4.36 \\
\textbf{95\% CI}     & 0.89--5.57 & 0.77--6.41 & 0.54--4.47 & 1.12--7.23 \\
\bottomrule
\end{tabular}}
\end{table}

Similarly, Table~\ref{tab:boot2} shows the bootstrap confidence intervals computed for each test condition in Table~IV. In this case, the $95$\% CI includes the null hypothesis value, indicating that there is no substantial difference ($p > 0.05$). Any-to-one and any-to-any mappings achieve a comparable level of de-identification performance, consistent with the pinhole effect.

\begin{table}[t]
\centering
\caption{Bootstrap-estimated 95\% confidence intervals (CIs) of EER differences ($d_{\text{EER}}$) for de-identification performance.}
\label{tab:boot2}
\resizebox{\columnwidth}{!}{
\begin{tabular}{lcccc}
\toprule
\textbf{System} & \textbf{Libri-Dev (F)} & \textbf{Libri-Dev (M)} & \textbf{Libri-Test (F)} & \textbf{Libri-Test (M)} \\
\midrule
\textbf{SYS1 (Mean)} & 0.41 & -0.38 & -0.89 & 2.43 \\
\textbf{95\% CI}     & -1.34--2.74 & -2.84--2.69 & -4.04--3.34 & -1.34--6.44 \\
\textbf{SYS2 (Mean)} & 0.36 & -0.14 & 1.54 & 0.61 \\
\textbf{95\% CI}     & -2.78--3.67 & -2.05--1.68 & -2.74--5.81 & -2.36--4.71 \\
\bottomrule
\end{tabular}}
\end{table}

\section{Conclusion}
We have introduced the \emph{pinhole effect} as a conceptual framework to explain the identity linkability behavior frequently observed in speaker anonymization systems. By modeling the anonymization process as a mapping function from original speaker identities to pseudo speakers, we examined two key strategies: mapping to a common pseudo speaker (any-to-one) and mapping to distinct pseudo speakers (any-to-any). Our analysis shows that anonymizing each utterance to a unique pseudo speaker significantly reduces speaker linkability by increasing the dispersion in the anonymized speaker space. While both mapping strategies achieve comparable de-identification performance (i.e, the anonymized speech cannot be reliably traced back to the original speaker), the use of unique pseudo speaker offers a clear advantage in lowering linkability, which is a desirable property in privacy-preserving speech processing. 


\appendix[Statistical Significance Test Using Bootstrap]

\noindent To assess whether the EER difference between the two methods (a2a and a2o) is statistically significant, a bootstrap procedure was employed as follows.

\begin{enumerate}
    \item \textbf{Generate Bootstrap Samples:}  
    From the original ASV score set of size $n$, multiple bootstrap samples were created by resampling with replacement. Each bootstrap sample contained $n$ data points, allowing duplicates; hence, some scores from the original dataset could appear multiple times, while others might not appear at all.

    \item \textbf{Compute the Statistic:}  
    For each bootstrap sample, the equal error rates (EERs) of both methods (a2a and a2o) were computed, and their difference was calculated as
    \[
        d_{\text{EER}} = \text{EER}_{\text{a2a}} - \text{EER}_{\text{a2o}}.
    \]

    \item \textbf{Build the Bootstrap Distribution:}  
    The resampling and computation process was repeated $1000$ times to obtain a bootstrap distribution of $d_{\text{EER}}$.

    \item \textbf{Determine the Confidence Interval:}  
    The bootstrap differences were sorted, and the $2.5$th and $97.5$th percentiles were taken to form the $95$\% confidence interval (CI).

    \item \textbf{Test for Statistical Significance:}  
    If the null hypothesis value (zero difference) lay outside the $95$\% CI, the null hypothesis was rejected, indicating that the EER difference between the two methods was statistically significant.
\end{enumerate}

\newpage

\bibliographystyle{IEEEtran}
\bibliography{ref}

\end{document}